\documentclass[sigconf]{acmart}

\PassOptionsToPackage{numbers, compress}{natbib}

\usepackage[utf8]{inputenc} 
\usepackage[T1]{fontenc}    
\usepackage{hyperref}       
\usepackage{url}            
\usepackage{booktabs}       
\usepackage{amsfonts}       
\usepackage{nicefrac}       
\usepackage{microtype}      
\usepackage{xcolor}         
\usepackage{graphicx}
\usepackage{subfig}
\usepackage{makecell}
\usepackage{amsmath}
\usepackage{tcolorbox}
\usepackage{wrapfig}

\newcommand{\nameee}{{\sc C2Rust-Bench}}
\newcommand{\fullset}{{15,503}}
\newcommand{\microset}{{1,573}}
\newcommand{\selectedset}{{2,905}}

\renewcommand\footnotetextcopyrightpermission[1]{}
\title{\nameee{}: A Minimized, Representative Dataset for C-to-Rust Transpilation Evaluation}

\author{Melih Sirlanci}
\affiliation{%
  \institution{The Ohio State University}
  \country{}
}
\email{sirlanci.2@osu.edu}

\author{Carter Yagemann}
\affiliation{%
  \institution{The Ohio State University}
  \country{}
}
\email{yagemann.1@osu.edu}

\author{Zhiqiang Lin}
\affiliation{%
  \institution{The Ohio State University}
  \country{}
}
\email{zlin@cse.ohio-state.edu}

\begin{document}

\renewcommand{\shortauthors}{Sirlanci, et al.}

\acmConference[Conference acronym 'XX]{}{}

\begin{abstract}
Despite the effort in vulnerability detection over the last two decades, memory safety vulnerabilities continue to be a critical problem. Recent reports suggest that the key solution is to migrate to memory-safe languages. To this end, C-to-Rust transpilation becomes popular to resolve memory-safety issues in C programs. Recent works propose C-to-Rust transpilation frameworks; however, a comprehensive evaluation dataset is missing. Although one solution is to put together a large enough dataset, this increases the analysis time in automated frameworks as well as in manual efforts for some cases. In this work, we build a method to select functions from a large set to construct a minimized yet representative dataset to evaluate the C-to-Rust transpilation. We propose \nameee{} that contains \selectedset{} functions, which are representative of C-to-Rust transpilation, selected from \fullset{} functions of real-world programs.
\end{abstract}

\maketitle

\section{Introduction}
The recent advancements in Artificial Intelligence (AI) bring back the discussion of code migration "transpilation" from one programming language to another. Certainly, transpilation of C programs into Rust is one of the most popular, since Rust provides memory-safety as opposed to the memory-unsafe C language while maintaining comparable runtime performance. It is reported that 70\% of the Common Vulnerabilities and Exposures (CVEs) assigned by Microsoft are about memory safety vulnerabilities~\cite{cisareport2023}. Instead of detection and mitigation of memory-safety vulnerabilities, the report from the US White House Office of National Cyber Directory (ONCD) identifies migration to memory-safe languages as the key solution~\cite{whitehousereport2024}. The program announced by the Defense Advanced Research Projects Agency (DARPA) aims to automate the transpilation of legacy C code into Rust~\cite{darpatractor}.

Initial attempts have been made to propose automated tools to translate C programs into Rust using Large Language Models (LLMs)~\cite{emre2021translating, zhang2023ownership, yang2024vert, hong2024don, shiraishi2024context}. However, there is a non-trivial question: \textit{What dataset should be used to evaluate the proposed tools?} As observed in existing program analysis works, such as vulnerability detection, an evaluation dataset should have two main features: concise and representative. The dataset has to be concise with a minimal set of samples to reduce and optimize the time spent evaluating the proposed tools. More importantly, a representative dataset provides a basis for evaluating tools with varying distinctive samples from a large comprehensive set. Testing transpilation tools is subject to such challenges, as transpilation can become increasingly costly and time-consuming depending on the underlying LLM while requiring a comprehensive testing.

To the best of our knowledge, there is no prior work that provides a representative program set for C-to-Rust transpilation task. However, existing works in other areas propose datasets with the desired two features, representative and concise. Existing work on vulnerability detection aims to build a representative dataset that can be used to evaluate vulnerability detection frameworks~\cite{dolan2016lava, hazimeh2020magma}. However, the proposed approaches cannot be used for transpilation, since they are domain-specific and do not aim to produce a concise dataset. Moreover, existing work in Machine Learning (ML) attempts to reduce the size of the training dataset to increase efficiency \cite{bachem2017practical, sener2017active, novikov2021dataset, lee2024coreset, song2025leave}. Although they share a similar motivation, they tackle a different problem, focusing on reduction of training data that often involves general-purpose data, such as text or images, with different characteristics than code written in a programming language.

Reducing a large dataset containing real-world C programs without losing any of the distinct program samples has several challenges. Programming constructs in a programming language can be thought of as small in number; however, the combinations of those constructs produce unique programs with varying complexity levels. This poses the first challenge as having a close to infinite number of programs to select a representative set from. Moreover, even with a finite program set, it would still be non-trivial to select representative programs, since it is challenging to describe source code in terms of the target task "transpilation". This requires an answer to the questions: \textit{What is a representative C program set for transpilation? How can we describe it?}

Even though combinations of programming constructs can create an infinite number of programs, based on our observations on LLMs and rule-based transpilation, they work on smaller units from input code at a time. This means that combining programming constructs and creating long programs (conceptually can be thought of as programs with infinite source lines of code) do not increase the complexity for transpilation linearly, nor affect representativeness. Moreover, representativeness can be defined as having a subset of programs that contains all challenging code pieces for transpilation coming from the large set. Based on this definition and our observations, we aim to obtain a set of metrics that express the complexity of a given program, which overcomes the challenge of lack of quantitative metrics for source code.

Based on our insights, we propose a method that takes a large dataset from real-world programs of various domains and selects programs with varying levels of representativeness based on complexity metrics. In the literature, there are various software metrics that represent different features of a given code. We find that the Maintainability Index (MI) suits our purpose well~\cite{coleman1994using} since it is the summary of three other metrics, including (1) cyclomatic complexity~\cite{mccabe1976complexity}, representing control flow complexity, (2) Halstead's volume~\cite{halstead1977elements}, representing the amount of information contained within a code, and (3) source lines of code, representing the size of the code. Since our target task is C-to-Rust transpilation and two languages differ in several aspects such as memory operations and data types, taking only C code into account for complexity measurement would be limited in expressing representativeness. Thus, in addition to MI of C code, we cover the MI of Rust code as a separate complexity metric. Moreover, we observed that it is challenging to transpile C code containing pointer arithmetics, memory operations, and advanced data structures. Based on those observations, we identify two other sources that represent the complexity of a Rust code for the transpilation task, namely the usage of unsafe code and the usage of varying data types.

In the selection process, we utilize partitioning by cutting each metric into pieces whose combinations form bins in multidimensional space. Then, we calculate the principal component analysis (PCA) complexity score, which is the summary of four metrics, and order each bin by the PCA complexity score. Lastly, we perform selection using systematic sampling from each bin to obtain representative samples from distinctive data points.  

To form an initial large dataset, we obtain the programs used in the evaluation of the previous C-to-Rust transpilation works~\cite{emre2021translating, zhang2023ownership, yang2024vert, hong2024don}. In the large dataset, there are 65 programs containing \fullset{} functions in total. Applying our selection process, we obtain \selectedset{} number of functions that form \nameee{}, selected by 81.3\% reduction of the large set. We release \nameee{} to be used in the evaluation of the C-to-Rust transpilation works~\footnote{https://doi.org/10.5281/zenodo.15249150}. In addition, we publish the code artifacts of our implementation on GitHub~\footnote{https://github.com/sirlanci/FuncSelector}.

\section{Overview}
In this section, we first give a background on dataset usage and potential challenges related to dataset in program analysis and machine learning. Then, we provide a brief background on the transpilation from C to Rust. Next, we explain our motivation behind the dataset reduction for the C-to-Rust transpilation.

\subsection{Background}
\paragraph{Dataset in program analysis.} In program analysis, one of the main focuses is to propose a tool for purposes such as vulnerability detection, malware detection, or transpilation. This requires those works to evaluate the proposed tools on a set of input programs (e.g., programs with CVEs, malware samples, or source code with different characteristics). In such cases, there are two main approaches to obtain a dataset, (1) picking a set of real-world programs from varying domains~\cite{emre2021translating} or (2) retrieving the dataset from previous work if it is available~\cite{zhang2023ownership}. Furthermore, some works collect real-world programs in addition to retrieving the existing dataset from previous works~\cite{hong2024don}. However, these approaches face potential problems: (1) having a small dataset leads to not evaluating the proposed tool comprehensively, (2) having a large dataset leads to a long analysis time, or (3) having a different dataset from previous work makes them incomparable. This is an optimization problem requiring one to find a representative smallest input set containing distinct samples.

\paragraph{Dataset in Machine Learning.} ML models are built from data through training and testing processes. Due to the dependency on data, dataset collection and selection have a critical role in ML works, which deals with the trade-off between the accuracy of the model and the training and testing time. To improve accuracy, a common way is to increase the amount of data, which requires more time to train and test the model. However, while increasing the amount of data, it is likely to increase the duplicated data, which does not have an impact on improving accuracy. Consequently, there are works on the reduction of dataset size in the ML literature~\cite{bachem2017practical, sener2017active, novikov2021dataset, lee2024coreset, song2025leave}. The dataset reduction works focus on obtaining a subset, which contains distinct or representative data points, from a large dataset. This reduction in the size of the dataset helps to reduce training and testing time while keeping accuracy at the same level.

\paragraph{Transpilation.} 
C-to-Rust transpilation gained popularity, as the migration of programs from the memory-unsafe to the memory-safe language has been seen as a permanent solution to memory-safety vulnerabilities~\cite{whitehousereport2024, darpatractor}. Initial effort has been put on rule-based C-to-Rust transpilation. Among rule-based transpilation tools, c2rust~\cite{immunantc2rust} achieves good results, producing compilable and semantically correct Rust code; however, the output is not idiomatic and is wrapped in unsafe blocks by default. Several works focus on improving the c2rust output by reducing unsafe usage~\cite{emre2021translating, zhang2023ownership} and modifying function returns with proper Rust types~\cite{hong2024don}. Despite improvement, rule-based transpilation is fundamentally limited. Thus, the focus has recently shifted to AI-based transpilation. A recent work proposes an LLM-based C-to-Rust transpiler, VERT~\cite{yang2024vert}, while another focuses on overcoming the context window limits in LLM transpilation~\cite{shiraishi2024context}.

\subsection{Motivation}
Although significant effort has been put into vulnerability detection~\cite{song2019sok}, memory safety vulnerabilities remain a critical threat~\cite{cisareport2023, whitehousereport2024}. Instead of putting effort into securing programs written in memory-unsafe languages, migrating programs to memory-safe languages, called transpilation, is considered the best alternative solution~\cite{whitehousereport2024, darpatractor}. 

One popular transpilation is from the C language to the Rust language. By design, the C language does not provide memory safety, intentionally supporting memory operations for system programming. However, Rust language provides a memory-safe alternative to C language, still allowing memory-unsafe operations through the \texttt{unsafe} keyword. Thus, the C-to-Rust transpilation becomes critical to implement the solution of migration from C to a memory-safe language~\cite{whitehousereport2024, darpatractor}.

Several works have been done on the C-to-Rust transpilation based on rule-based approach~\cite{emre2021translating, zhang2023ownership, hong2024don}, and LLM-based approach~\cite{yang2024vert, shiraishi2024context}, marking the inception of a new research area. However, as also observed in other fields under program analysis, it is challenging to determine a dataset to evaluate the proposed tools, due to potential problems: (1) having a small dataset that leads to not evaluating the proposed tool comprehensively, (2) having a large dataset that leads to a long analysis time, automated as well as manual analysis in some cases, or (3) having a different dataset from previous work that makes them incomparable. 

To avoid such issues in this newly emerging field of research, we form a benchmark set \nameee{}, containing \selectedset{} distinct and representative functions selected from \fullset{} functions in the large set by reducing it by 81.3\%. \nameee{} avoids the three potential problems previously mentioned by providing (1) a small but representative dataset that allows comprehensive evaluation, (2) a minimized dataset that reduces the analysis time by approximately 80\%, and (3) a standardized dataset that offers a common ground for the comparison of incremental works.

\section{Methodology}
In this section, we first present an overview of our methodology. Then, we explain the source code complexity metrics. Lastly, we present the function selection method. 

\subsection{Overview}
Selection of representative functions from a large dataset for the target task, which is evaluation of C-to-Rust transpilation, requires (1) identifying features that represent the functions with respect to the target task and (2) a method to select functions based on the features. We identify and define a set of source code complexity metrics for C and Rust. To obtain Rust code corresponding to an input C code, we build a transpilation framework that uses a local LLM. Consequently, this allows us to get features from both sides, input and output, of the C-to-Rust transpilation. We present the implications of using an LLM to construct an evaluation dataset for LLM transpilation in~\S\ref{sec:cross_evaluation}.

Selecting functions based on code complexity metrics requires a method to find distinctive data points in a multidimensional space. We utilize a partitioning method that cuts each dimension into pieces based on a specified partitioning parameter. Then, the combinations of those pieces in the multidimensional space create bins. Lastly, we order samples in each bin using the Principal Component Analysis (PCA) score calculated from the code metrics of each sample and perform systematic sampling to select representative and distinctive functions from each bin. In the following subsections, we present our methodology in detail.

\subsection{Representative Source Code Complexity Metrics}
\label{sec:code_metrics_methodology}
In this subsection, we explain the source code metrics that we obtained or defined to represent the complexity of functions to be used in the selection process. 

\paragraph{Maintainability Index.} One popular metric to measure the complexity of a source code is the Maintainability Index (MI) \cite{oman1992metrics}. The MI metric is built on top of three other metrics, cyclomatic complexity~\cite{mccabe1976complexity}, Halstead's volume~\cite{halstead1977elements}, and source line of code. Those three metrics are combined in a formula to get different aspects of the code involved in the measurement of the maintainability of a code. 

MI metric shows how easy it is for developers to understand, modify, and extend a piece of code. Taking into account the underlying three metrics,  MI can be considered as a complexity metric that shows the complexity of the code in the development process. The transpilation task requires understanding the original code and rewriting it in another language without changing the functionality. From this aspect, transpilation can be thought of as a part of the development process. This makes the MI metric an excellent candidate for being used in the representative code selection process.

\paragraph{Unsafe Code Complexity.} Memory operations, such as pointer arithmetic and dereferencing, present a significant challenge in the transpilation from C to Rust due to fundamental differences in how memory safety is handled in these languages. In C, these operations are freely allowed without any safety checks, placing the responsibility on the programmer to manage memory correctly. However, Rust enforces strict memory safety guarantees, ensuring that operations such as pointer manipulation are performed only in contexts where the programmer can guarantee their safety. Transpilation in such a case is challenging because it requires the transpiler to account for Rust's memory safety rules in producing code.

Considering previous points and our observations using LLM for transpilation, we believe that the use of unsafe code in Rust increases the complexity of a code in terms of transpilation. Since the MI index does not take memory operations into account while measuring maintainability, we define a metric to represent the complexity of a Rust code with respect to its unsafe operations. We obtain unsafe blocks and count the number of unsafe statements. Then, we find the average unsafe statement for a given Rust function. The details of the implementation of this metric are presented in~\S\ref{sec:data_collection_implementation}.

\paragraph{Data Type Complexity.} Data types present a significant challenge in the transpilation process from C to Rust due to fundamental differences in type systems, memory layouts, and handling of basic and complex data types between the two languages. In C, the type system is relatively permissive, allowing implicit type conversions, type casts, and low-level memory manipulations. For instance, C allows one to cast between incompatible types and perform arithmetic on raw pointers, which can lead to memory misalignments or violations of type safety. In contrast, Rust enforces a stricter type system that prioritizes safety and explicitness. Rust's type system disallows implicit type coercion and enforces explicit casting between types requiring careful handling during transpilation. Furthermore, Rust introduces advanced concepts such as ownership, borrowing, and lifetimes that create complexities in transpilation, as they introduce additional constraints related to memory allocation and ownership that are absent in C. 

Translating data types from C to Rust requires not only mapping flexible memory handling of C language to Rust's more restrictive model but also ensuring that the resulting code respects Rust's memory safety, ownership, and lifetime rules. The challenge is to correctly map both basic and complex data types with necessary conversions and safety checks to conform to Rust's type system and memory management practices. It is evident that translating data types from C to Rust is a challenging process. Thus, we include a metric that reflects the complexity of a given Rust program based on the data types used in a given Rust function. The details of the implementation of this metric are shared in~\S\ref{sec:data_collection_implementation}.

\subsection{Function Selection Method}
\label{sec:func_select_methodology}
We obtain and define 4 metrics to quantify the complexity of C and corresponding Rust functions, which include the MI of C code, the MI of Rust code, unsafe code complexity metric of Rust code and data type complexity metric of Rust code. To facilitate function selection, we require a method that allows us to collectively analyze these four metrics and identify representative data points that capture the characteristics of the entire set of functions. k-means clustering appears to be a suitable approach, as it can categorize data based on multiple features. However, we observe that the values of our complexity metrics are continuous without having any gaps to form groups, which makes clustering algorithms infeasible to utilize for our purpose.

For data with continuous values, one practice is to create subgroups by dividing the population into subgroups based on some characteristics, as in stratified sampling. Since we do not have such categorical characteristics, we use an alternative but similar approach, partitioning, to create subgroups. Basically, we partition the value range of each metric by cutting it into equal pieces. In a multi-dimensional space, a combination of those pieces from each dimension create bins, which are similar to naturally formed subgroups in stratified sampling. Each of these bins represents a distinctive data point in the multidimensional space. 

Since we aim to obtain a representative set with distinct samples, we employ systematic sampling and Principal Component Analysis (PCA) to use in the selection from each bin. First, we calculate PCA for the samples in each bin, which summarizes the four metrics that produce one single complexity metric. Next, we order samples in each bin using the PCA metric. Then, we perform systematic sampling from each bin based on the interval value calculated from sampling size of each bin. The details of the implementation of our function selection process are presented in~\S\ref{sec:function_selection_implementation}.

\section{Implementation}
In this section, we present the implementation of the components of our selection process. In~\S\ref{sec:preparation_implementation}, we share how the C files are preprocessed to preprare for transpilation. Then, we explain the transpilation of C functions into Rust in~\S\ref{sec:transpilation_implementation}. Lastly, we present the implementation of complexity metrics and function selection method in~\S\ref{sec:data_collection_implementation} and~\S\ref{sec:function_selection_implementation}, respectively. 
\subsection{Preparation of C Code for Transpilation}
\label{sec:preparation_implementation}
\paragraph{Preprocessing C Files.} At this first step, we perform preprocessing on C files with the aim of resolving macros and dependencies among C files in a codebase. This preprocessing step is required to reduce the input from multiple C files to single C file, since currently transpilation using LLMs faces challenges such as the input and output token limit.

We modify the compilation files of each codebase in our dataset to produce preprocessed C files. We update the build files such as \texttt{Makefile} with the \texttt{-S} option of GCC, which produces a preprocessed C file instead of a compiled binary file. Those preprocessed C files can be individually further compiled into a binary file. Thus, we work on those preprocessed C files in the segmentation process.

\paragraph{Segmentation of C Files.} Two challenges force us to split a C file into individual functions, including (1) LLMs have input and output token limits, and (2) while it is challenging to produce compilable transpilation output using LLMs for individual functions, it becomes nearly impossible to do the same for a C file with multiple functions. Although the topic has recently become popular, there is one work that shows the effectiveness of segmentation in the C-to-Rust transpilation \cite{shiraishi2024context}. Unfortunately, the source code of the proposed segmentation approach is not publicly available.

We build an LLVM tool to split a preprocessed C file into individual functions. The tool first identifies the start and end lines of each function in a given C file. Then, it extracts the functions into individual C files. The source code of the LLVM tool can also be found in our artifacts.

\subsection{Transpilation of C Functions into Rust}
\label{sec:transpilation_implementation}
We need a method to perform transpilation from C to Rust since we need, (1) to obtain the complexity metrics from the transpiled Rust code, and (2) to get feedback from the transpilation process to use in the evaluation of the selected functions. To this end, we build a simple transpilation tool with a compilation error fixing loop, which can work with any local LLM. We explain our choice of the LLM used in the experiments in \S\ref{sec:llm_selection}. 

The tool gets a C file that contains an individual function as an input. Then, it combines the input C function with the instructions shown in \autoref{fig:transpile_instruction} of Appendix \ref{appendix:transpilation_implementation}, and sends a request to the LLM. The LLM performs transpilation and returns the transpiled Rust code in the format specified by the instructions. Then, the tool attempts to compile the Rust code. If the compilation is successful, the transpiled Rust code is kept, and the transpilation process ends with success. Otherwise, the tool obtains the compilation errors generated by the Rust compiler and passes them to the fixing module along with the corresponding Rust code. 

The fixing module sends a request to LLM including the combination of the Rust code, compilation errors, and fixing instructions that are shown in~\autoref{fig:fixing_instruction} of Appendix \ref{appendix:transpilation_implementation}. Then, the fixing module tries to compile the Rust code that is modified to fix compilation errors by LLM. Until the compilation error fixing attempt limit is reached, the fixing module attempts to fix the transpiled Rust code. If the transpiled Rust code is not compilable when the limit is reached, the transpilation process ends with failure. Although not compilable, we keep those samples in our selection candidate set, since they are evidently challenging for transpilation, and complexity metrics can still be obtained for them.  

\subsection{Data Collection of Code Metrics}
\label{sec:data_collection_implementation}
In this subsection, we present how we implement the data collection of the code complexity metrics. The details of how we utilize existing tools and build parsers to get complexity metrics can be found in the GitHub~\footnote{https://github.com/sirlanci/FuncSelector}.

\paragraph{Maintainability Index}
There are tools to measure Maintainability Index (MI) of a code from C and Rust languages. We use the open-source tool published on github to obtain MI metric for a C function~\cite{Jarod42}. We employ the rust code analysis tool published in a previous work to obtain MI metric for the corresponding Rust code~\cite{ardito2020rust}. 

\paragraph{Unsafe Operation Complexity}
\label{sec:unsafe_blocks}
We implement unsafe operation complexity metric by building a parser in Rust, specifically designed to identify and extract unsafe blocks from a given Rust function. One important detail is that we choose to count unsafe blocks, as well as statements inside the blocks, because counting only blocks would be limited to express the complexity of unsafe operations, as unsafe blocks may contain a variable number of statements, ranging from a single operation to a large number of operations.

As a result, the parser produces two key outputs: the total number of unsafe blocks and the set of number of statements contained in each block. To represent unsafe operations as a unified metric, we calculate the average number of unsafe statements per unsafe block. Using this metric, we capture not only the frequency of unsafe code usage but also its density. The lower boundary of this metric is zero, indicating the absence of unsafe code, while the upper boundary is unbounded, reflecting the potential for increasingly complex unsafe operations.

\paragraph{Data Type Complexity}
\label{sec:unique_data_types}
In Rust, variables can be explicitly typed, or the type may be inferred by the compiler based on the value assigned. To quantify the complexity introduced by data types, we build a parser to collect and analyze the types of variables within a given Rust code. The parser extracts the type of a variable when it is explicitly specified in the declaration. For example, when a variable is defined as \texttt{let x:i32=5}, the parser identifies the type as \texttt{i32} for the variable \texttt{x}. 

However, unlike languages such as C, Rust allows variables to be defined without explicitly stating their type. In these cases, the parser examines the right-hand side of the declaration, if available, and infers the variable’s type based on the right-hand side expression. For example, in a declaration such as \texttt{let y=10}, the parser infers that the type of variable \texttt{y} is \texttt{i32}, since the value of \texttt{10} is an integer literal.

After extracting all the variable types, we get the set of unique types and obtain the total number of types in the set. We use the number of unique types as a metric to represent the complexity of the data types for a given Rust function.

\subsection{Function Selection using Partitioning}
\label{sec:function_selection_implementation}
As a first step in obtaining bins, we divide each of the dimensions, corresponding to the complexity metrics, into partitions based on the value of the width specified by the following formula:

\begin{equation}
\textit{width} = \frac{\textit{max\_value} - \textit{min\_value}}{\textit{number\_of\_partition}}
\label{eq:width}
\end{equation}

The $number\_of\_partition$ used in \autoref{eq:width} is a hyperparameter common for all dimensions, specifying the number of pieces to cut. We present a preliminary experiment identifying the optimal value of $number\_of\_partition$ in \S\ref{sec:tuning_hyperparameters}. The width value in \autoref{eq:width} is calculated for each dimension, since the metrics have different minimum and maximum values. Based on the width values, each dimension is partitioned into equal pieces. Then, the combination of the partitions in multidimensional space forms the bins. If $number\_of\_partition$ is set to $n$, the number of bins created is equal to $n^4$. However, some of those bins remain empty as anticipated, and we proceed with the non-empty bins. 

We select samples from each bin, using Principal Component Analysis (PCA) and systematic sampling. We first calculate a summarized complexity metric from 4 metrics for each sample using PCA. Then, we order samples in each bin using the unified PCA complexity metric. Next, we perform selection from each bin based on the systematic sampling approach, which means that we select from ordered samples that are placed away from each other by an interval. In the following formula, we obtain the interval value to use in systematic sampling:

\begin{equation}
\textit{interval} = \frac{\textit{bin\_population}}{\textit{bin\_population} \times \textit{ratio\_of\_sampling}}
\label{eq:interval}
\end{equation}

The $ratio\_of\_sampling$ used in \autoref{eq:interval} is another hyperparameter common for all bins, which specifies the percentage of sample to take from each bin. The optimal value of $ratio\_of\_sampling$ parameter is specified by a preliminary experiment in \S\ref{sec:tuning_hyperparameters}. In the denominator of this formula, we have the sampling size calculated by multiplying $subset\_population$ by $ratio\_of\_sampling$, which is different for each bin. We obtain the interval value for each bin by dividing the population size by the sampling size of the bin. Lastly, we select samples from each bin that are placed at positions separated by the interval value. Consequently, we obtain diverse samples from distinct data points depicted by the bins in multidimensional space and the different positions in each bin.

\begin{table*}[h!]
\centering
\footnotesize
\setlength\tabcolsep{9pt}
\caption{The detailed information of the programs in the large dataset.}
\label{tab:fullset_info}
\begin{tabular}{lrr||lrr|lrr}
\toprule
\textbf{Name} & \textbf{Function (\#)} & \textbf{SloC (\#)} & \textbf{Name} & \textbf{Function (\#)} & \textbf{SloC (\#)} & \textbf{Name} & \textbf{Function (\#)} & \textbf{SloC (\#)} \\
\hline
\hline
transcoder-set & 4,012 & 17,684 &       json-c & 154 & 2,201 &  gzip-1.12 & 28 & 780 \\ 
libxml2 & 1,964 & 26,729 &      rcs-5.10.1 & 154 & 1,955 &      bzip2 & 22 & 625 \\ 
gprolog-1.5.0 & 758 & 8,119 &   bc-1.07.1 & 153 & 2,581 &       genann & 22 & 317 \\ 
nettle-3.9.1 & 646 & 8,103 &    uucp-1.07 & 152 & 4,863 &       snudown & 20 & 293 \\ 
json.h & 644 & 18,415 &         findutils-4.9.0 & 151 & 2,985 &         libcsv & 20 & 180 \\ 
tmux & 607 & 7,625 &    pth-2.0.7 & 137 & 2,016 &       quadtree-0.1.0 & 20 & 160 \\ 
tulipindicators-0.9.1 & 516 & 5,448 &   dap-3.10 & 129 & 3,942 &        which-2.21 & 18 & 458 \\ 
libosip2-5.3.1 & 515 & 5,005 &  cpio-2.14 & 118 & 1,536 &       sed-4.9 & 18 & 184 \\ 
tar-1.34 & 446 & 6,528 &        binn-3.0 & 116 & 961 &  urlparser & 18 & 178 \\ 
less-633 & 343 & 4,947 &        units-2.22 & 104 & 2,307 &      robotfindskitten & 12 & 158 \\ 
nano-7.2 & 324 & 4,368 &        enscript-1.6.6 & 78 & 2,273 &   avl & 8 & 62 \\ 
optipng-0.7.8 & 311 & 5,548 &   hello-2.12.1 & 77 & 979 &       rgba & 7 & 69 \\ 
gawk-5.2.2 & 307 & 7,135 &      ed-1.19 & 69 & 691 &    bst & 5 & 59 \\ 
mcsim-6.2.0 & 260 & 4,208 &     brotli-1.0.9 & 62 & 1,084 &     xzoom & 4 & 419 \\ 
heman & 257 & 3,179 &   lodepng & 59 & 542 &    ht & 4 & 38 \\ 
screen-4.9.0 & 229 & 5,680 &    pexec-1.0rc8 & 54 & 598 &       qsort & 3 & 27 \\ 
wget-1.21.4 & 226 & 3,338 &     diffutils-3.10 & 45 & 1,593 &   libtool-2.4.7 & 2 & 37 \\ 
tinycc & 207 & 2,666 &  lil & 40 & 312 &        grabc & 1 & 48 \\ 
patch-2.7.6 & 182 & 2,360 &     buffer-0.4.0 & 39 & 332 &       libtree-3.1.1 & 1 & 0 \\ 
cflow-1.7 & 181 & 2,604 &       grep-3.11 & 38 & 1,263 &        glpk-5.0 & 1 & 0 \\ 
mtools-4.0.43 & 172 & 2,118 &   libzahl-1.0 & 35 & 653 &        time-1.9 & 1 & 0 \\ 
make-4.4.1 & 168 & 3,597 &      indent-2.2.13 & 32 & 694 \\ 
\bottomrule
\end{tabular}
\end{table*}

\section{Experiments and Results}
In this section, we first explain the experiment setup in~\S\ref{sec:experiment_setup}. Then, we present a preliminary experiment in~\S\ref{sec:llm_selection}, determining an LLM to use in the transpilation from C to Rust. Next, in~\S\ref{sec:tuning_hyperparameters}, we present another preliminary experiment to specify the set of hyperparameters to be used in the selection process. 
In~\S\ref{sec:cross_evaluation}, we perform a cross-LLM evaluation on the microbenchmark set, showing that the selected functions using chosen LLM are also representative for other LLMs. 
Lastly, we present \nameee{} that is a minimized, representative set for evaluating C-to-Rust transpilation in~\S\ref{sec:final_selection}.

\subsection{Experiment Setup}
\label{sec:experiment_setup}
\paragraph{Dataset.} We assembled a large dataset by collecting datasets from previous works~\cite{emre2021translating, zhang2023ownership, yang2024vert, hong2024don}. These works focus on C-to-Rust transpilation or on improving transpiled Rust code. Note that our large dataset also contains c2rust examples~\cite{immunantc2rustexamples}, which are included in the four datasets. There are 64 real-world programs and 1 synthetic program set in our large dataset, all of which come from those 4 previous works. After preprocessing and segmentation steps, we obtained \fullset{} functions in total. In~\autoref{tab:fullset_info}, we share the details of 65 programs in our large set.

\paragraph{Microbenchmark Set.} We need to specify an LLM to transpile C code into Rust as a part of selection, however, transpiling the entire large dataset with all candidate LLMs would not be feasible considering time and resource requirements. To overcome such an issue, we sample \textasciitilde 10\% of the functions in the large dataset and obtain \microset{} functions that form the microbenchmark set to use in LLM selection.

\paragraph{Resource and Environment.} We run our experiments on a server with an Intel Xeon Silver 4310 CPU and an NVIDIA A30 GPU. The server runs on a Ubuntu 22.04.4 LTS OS. We implement the Rust parsers using the Rust parsing library Syn. We implement the rest of our method in Python. 

\subsection{Evaluation of LLMs on Microbenchmark Set}
\label{sec:llm_selection}
Our selection approach benefits from both the C and the corresponding transpiled Rust code. Thus, we need to perform transpilation as part of our selection process. Since transpilation of the large set with \fullset{} functions is costly to perform multiple LLMs, we specify an LLM to use in transpilation. However, we assess how well the selections obtained using the chosen LLM generalize to other LLMs in~\S\ref{sec:cross_evaluation}.

\paragraph{Methodology.} We utilize the transpilation process explained in \S\ref{sec:transpilation_implementation} to evaluate LLMs on microbenchmark set. We obtain a set of popular LLMs that satisfy requirements, (1) that run on local machine, (2) that have model size less than 24 GB, and (3) that are reported to be promising on code tasks. We obtained 9 candidate LLMs after excluding some of them, which are either small or old models, return output that is not in the format specified in our instructions (e.g., containing natural language content, or code split in multiple parts).

We run the transpilation process on each candidate LLM with the microbenchmark set as input. We collect several metrics during the transpilation process to use in choosing an LLM, which include the result of the transpilation process, the initial transpilation time, the number of compilation-fixing attempts, and the total fixing time.

\begin{table}[]
\centering
\footnotesize
\setlength\tabcolsep{6pt}
\caption{Transpilation performance of 9 LLMs on the microbenchmark set.}
\label{tab:llm_evaluation}
\begin{tabular}{lrrrrr}
\toprule
\textbf{LLM Name} & \textbf{\thead{Average \\Transpilation\\Time (sec)}} & \textbf{\thead{Average \\Compilation \\Attempt (\#)}} & \textbf{\thead{Transpilation \\Success (\%)}} \\
\hline
\hline
codegeex4:9b & 92.8 & 10.7 & 51.2 \\
codestral:22b & 94.8 & 4.5 & 92.7 \\
gemma2:9b & 104.9 & 12.4 & 47.0 \\
llama3.2:3b & 79.9 & 12.5 & 59.9 \\
llama3.1:8b & 99.5 & 8.9 & 80.5 \\
mistral:7b & 92.5 & 12.4 & 55.0 \\
qwen2.5-coder:7b & 79.7 & 10.0 & 60.1 \\
qwen2.5-coder:14b & \textbf{61.0} & 3.3 & 96.2 \\
qwen2.5-coder:32b & 103.6 & \textbf{2.6} & \textbf{97.6} \\
\bottomrule
\end{tabular}
\end{table}

\paragraph{Results.} In \autoref{tab:llm_evaluation}, we present the results of transpilation performed using 9 LLMs with varying model sizes. The total time taken for one sample contains the initial transpilation time, compilation error fixing times, and compilation times. The average time shows the amount of time spent for transpilation on average for the \microset{} samples of the microbenchmark set. The compilation attempt shows the number of compilation error fixing performed, recalling that the number of compilation fixing attempts is 0 if the Rust code is compiled right after initial transpilation while the upper limit is 20, where the transpilation process ends with failure. The average compilation attempt shows the number of fixing attempts performed on average for the \microset{} samples of the microbenchmark set. Lastly, the transpilation success shows the ratio of the number of samples for which a compilable transpiled Rust code is produced in at most 20 compilation error fixing attempts.

Among 9 LLMs, \texttt{qwen2.5-coder:14b} takes the least amount of time that is 61 seconds on average for transpilation. \texttt{llama3.2:3b} and \texttt{qwen2.5-coder:7b} have the third and second least time with 79.9 and 79.7 seconds, respectively. \texttt{qwen2.5-coder:32b} performs the least number of compilation error fixing attempts with 2.6 fixing on average, while it is followed by \texttt{qwen2.5-coder:14b} and \texttt{codestral:22b} with 3.3, and 4.5 fixing attempts on average. Lastly, \texttt{qwen2.5-coder:32b} performs best in terms of transpilation success, producing a compilable transpiled Rust code in 97.6\% of cases. \texttt{qwen2.5-coder:14b} and \texttt{codestral:22b} have 96.2\% and 92.7\% transpilation success rate, respectively.

We choose \texttt{qwen2.5-coder:32b} to use in the rest of our experiments since it is the most successful in the transpilation task by producing the highest compilable transpiled Rust code. Here, we consider the highest compilation rate in our decision, since some of the future work that will benefit from our dataset might require compilation of the transpiled Rust code.

\begin{figure}[]
\centering
\includegraphics[width=0.45\textwidth]{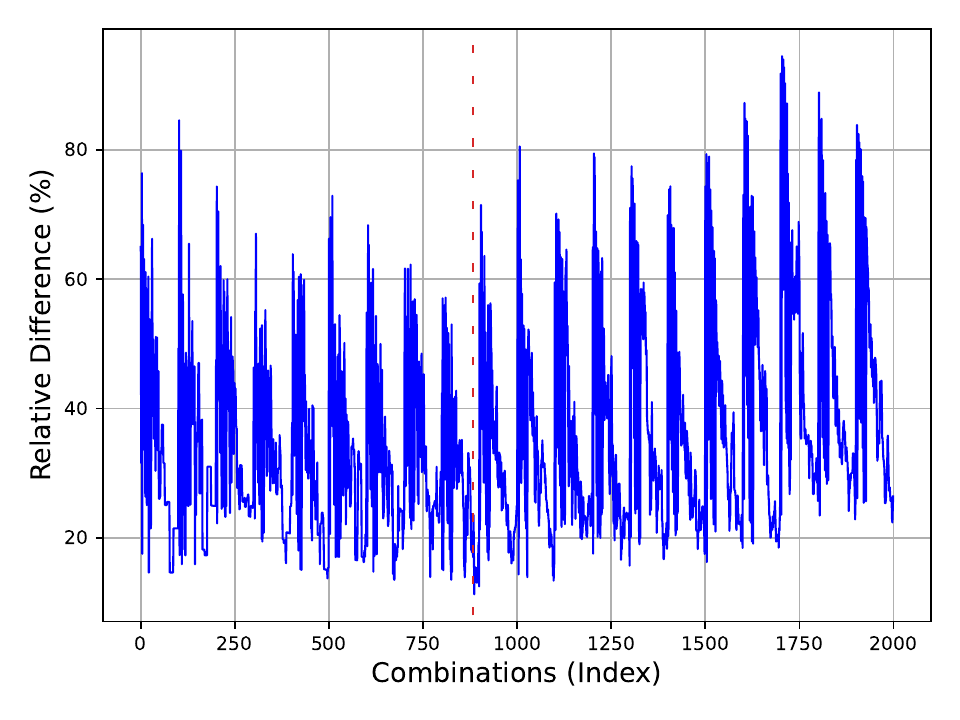}
\caption{The change of relative difference over combinations of values of two hyperparameters.\label{fig:squared_diff}}
\end{figure}

\subsection{Tuning the Hyperparameters}
\label{sec:tuning_hyperparameters}
The selection process has two hyperparameters, the number of partitions per dimension and the ratio of sampling per bin. In this section, we present the experiment performed to obtain the best set of hyperparameters. 

\paragraph{Methodology.} To find the best set of hyperparameters, we test different combinations of the two hyperparameters in the function selection process. We perform the selection following the process presented in~\S\ref{sec:data_collection_implementation} and~\S\ref{sec:function_selection_implementation}.

Evaluating the set of selections requires feedback from the target task that is transpilation in our case. We could use the transpilation result as successful and failed for feedback; however, it would give limited feedback since it is a binary result. Instead, we leverage the compilation-error fixing attempts from transpilation as feedback to justify the representativeness of the selections. The distributions of compilation error fixing attempts would be different for the large set and the selected function set since one is a subset of the other. To overcome such a difference, we normalize the compilation-error fixing attempt distribution of the large set by the ratio of number of samples in those two sets and obtain the expected distribution.

We calculate a difference score to justify the representativeness of the function sets selected with different hyperparameters. We use relative difference instead of absolute difference for a fair comparison, since the number of selected functions varies for each hyperparameter combination. We calculate the sum of differences from each compilation error fixing attempt and get the average. The calculation used for each combination of hyperparameters is shown in \autoref{eq:diff_score}.

\begin{equation}
\textit{relative\_difference} = \frac{1}{21} \cdot \sum_{i=0}^{20} \frac{|\textit{expected\_value}_i - \textit{observed\_value}_i|}{\textit{expected\_value}_i}.
\label{eq:diff_score}
\end{equation}

For the number of partition hyperparameter, we test 20 values in total, ranging from 1 to 20 and for the ratio of sampling hyperparameter, we test 100 values in total, ranging from 0.002 to 0.2 by 0.002 increase. Since these two hyperparameters are connected and work together, we tune them together by combining and testing their candidate values, which gives us a global optimum instead of a local optimum of each parameter. The combination of two hyperparameters produces 2,000 different selections in total.

\begin{figure}[h!]
\centering
\includegraphics[width=0.45\textwidth]{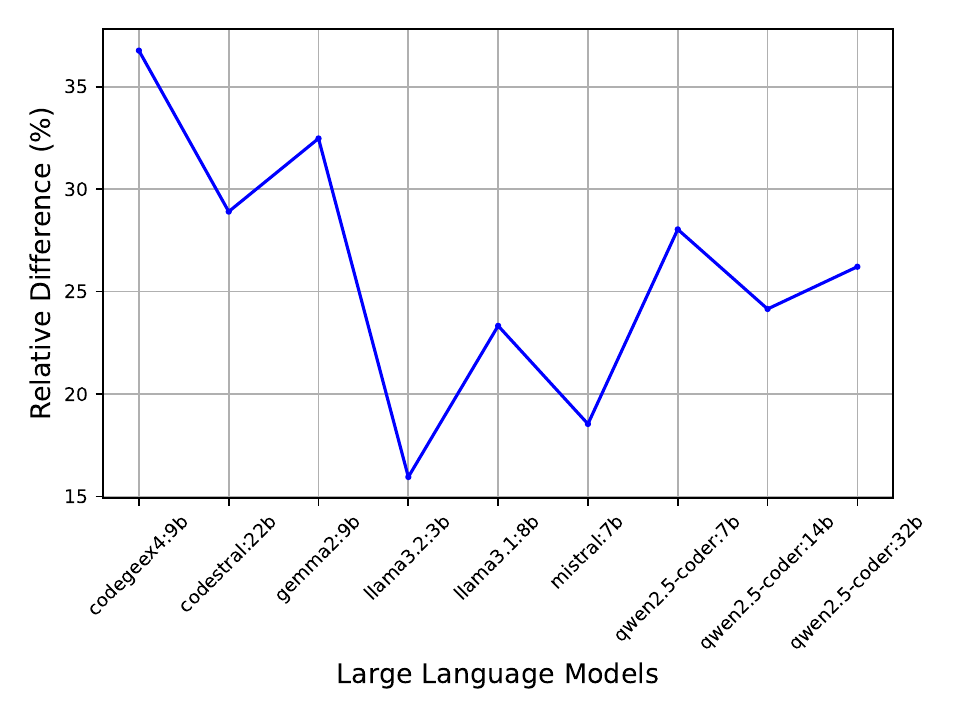}
\caption{The relative difference for 9 LLMs.\label{fig:llm_diff}}
\end{figure}

\paragraph{Results.} In \autoref{fig:squared_diff}, we present the relative differences calculated for 2,000 different selections from the combinations of two hyperparameters. The combinations of two hyperparameters are on the x-axis in the format of (number of partition, ratio of sampling). Each point on the x-axis corresponds to a unique combination of the two hyperparameters, ordered first by the value of the first hyperparameter, and then by the second. At the index 0 of the x-axis that is the leftmost in~\autoref{fig:squared_diff}, we have the combination (1, 0.002) and at the rightmost the combination (20, 0.2).

\begin{figure*}[h!]
    \centering
    \subfloat[codegeex4:9b]{\includegraphics[width=0.25\textwidth]{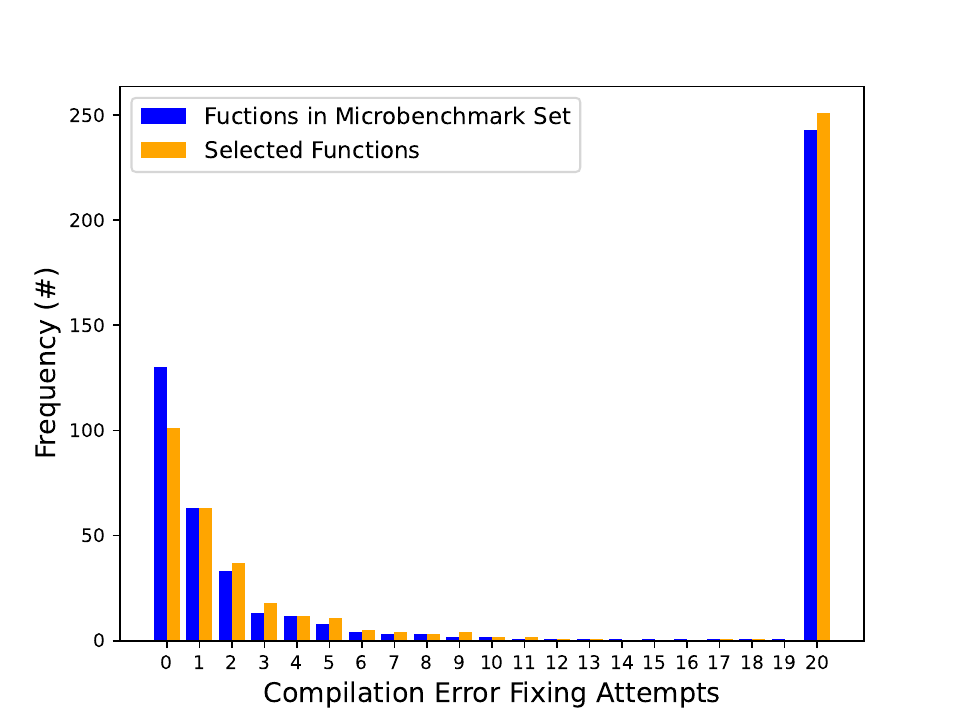}}%
    \subfloat[codestral:22b]{\includegraphics[width=0.25\textwidth]{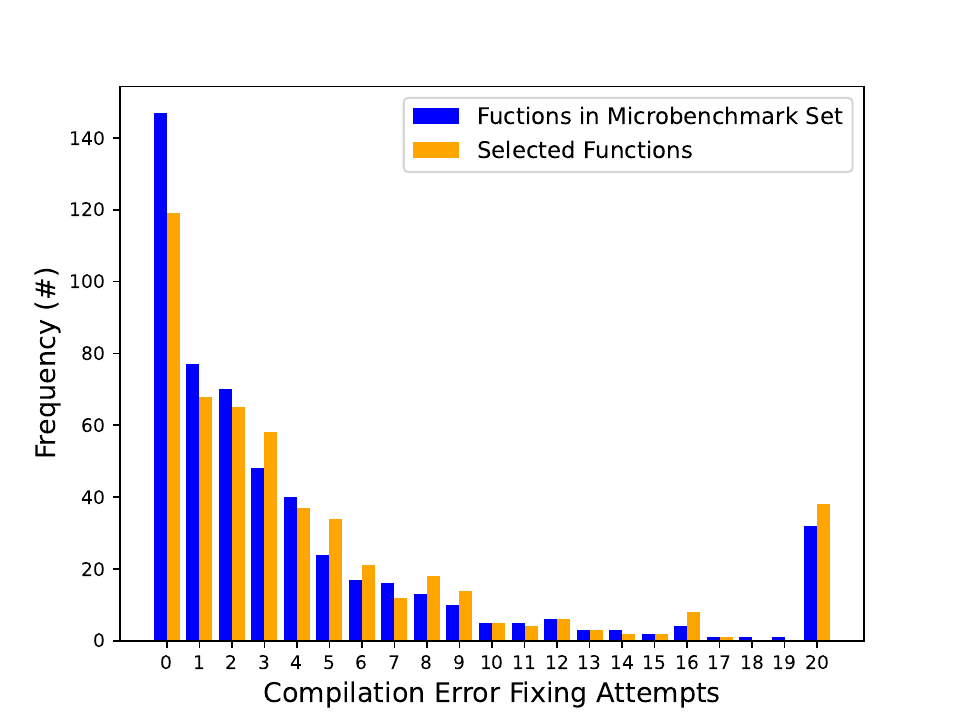}}%
    \subfloat[gemma2:9b]{\includegraphics[width=0.25\textwidth]{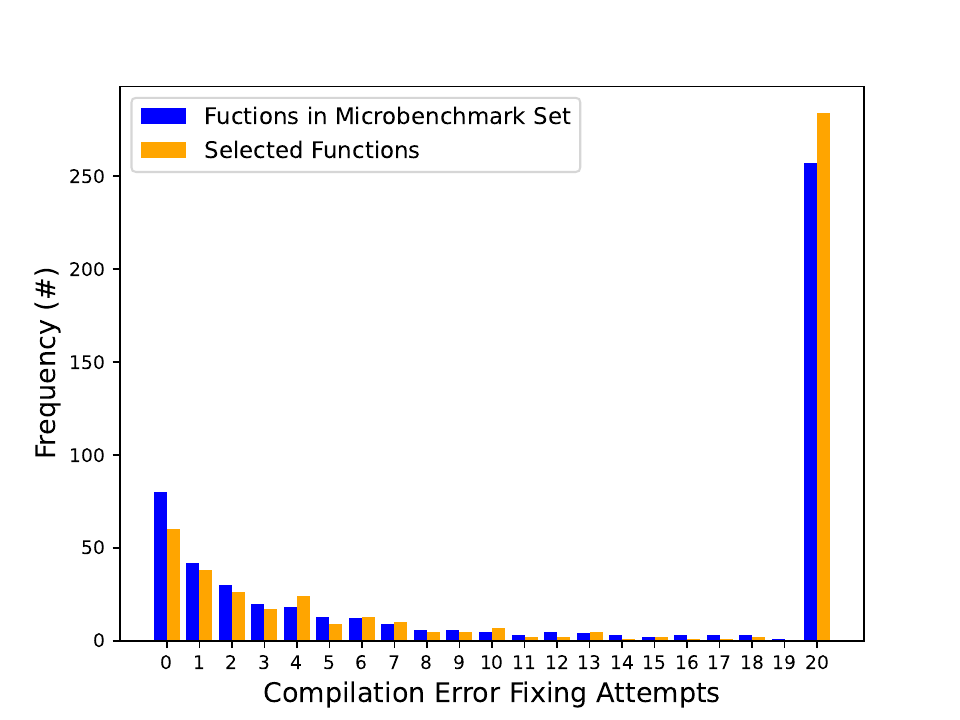}} \\[0.0cm]
    
    \subfloat[llama3.1:8b]{\includegraphics[width=0.25\textwidth]{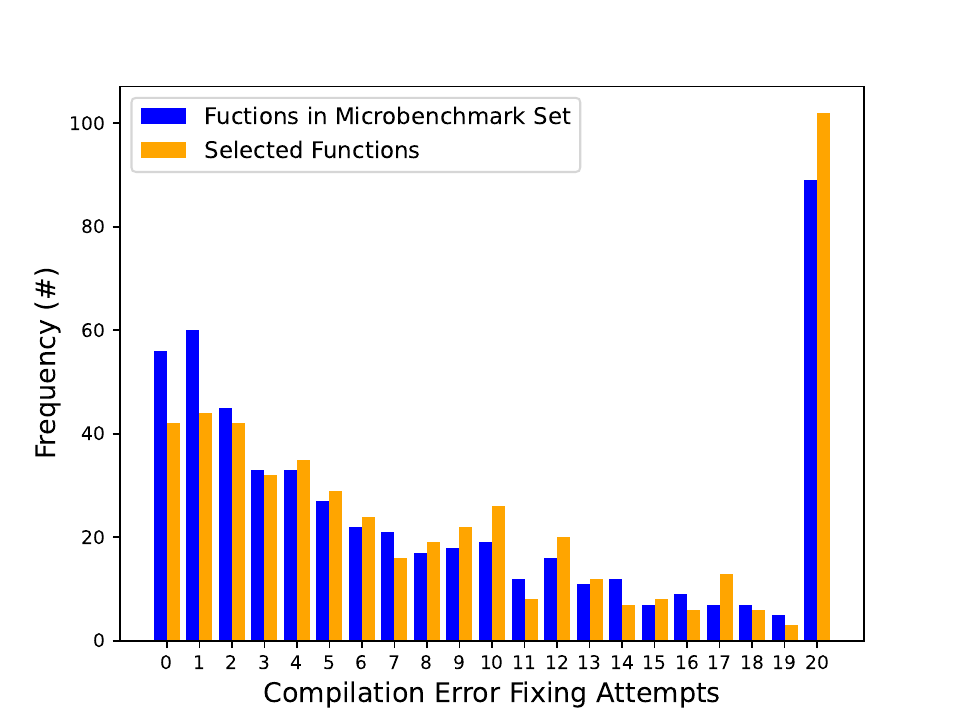}}%
    \subfloat[llama3.2:3b]{\includegraphics[width=0.25\textwidth]{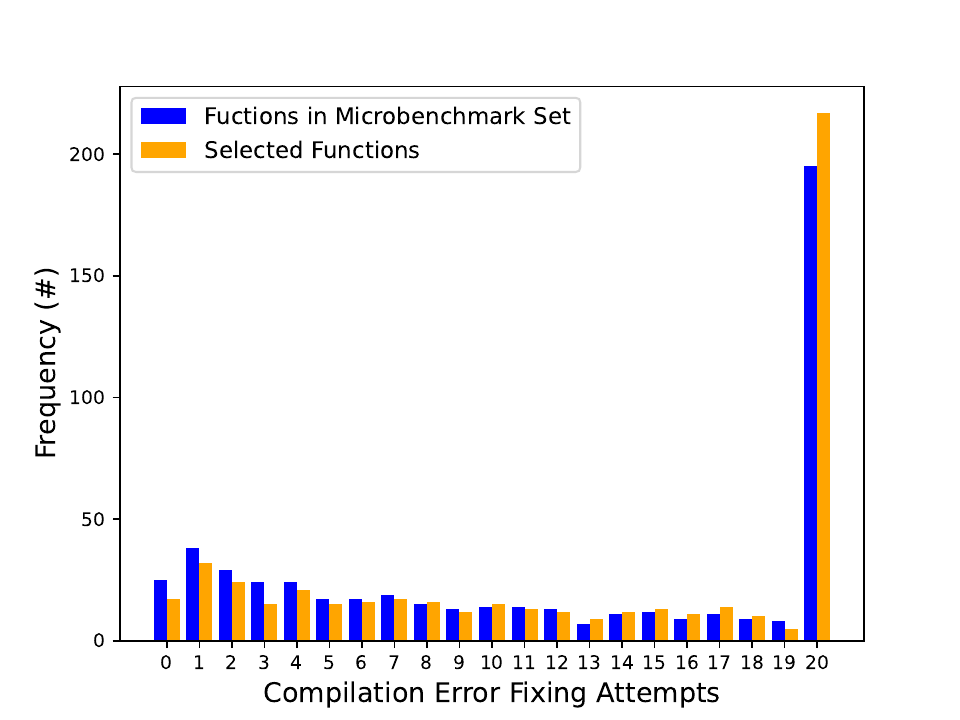}}%
    \subfloat[mistral:7b]{\includegraphics[width=0.25\textwidth]{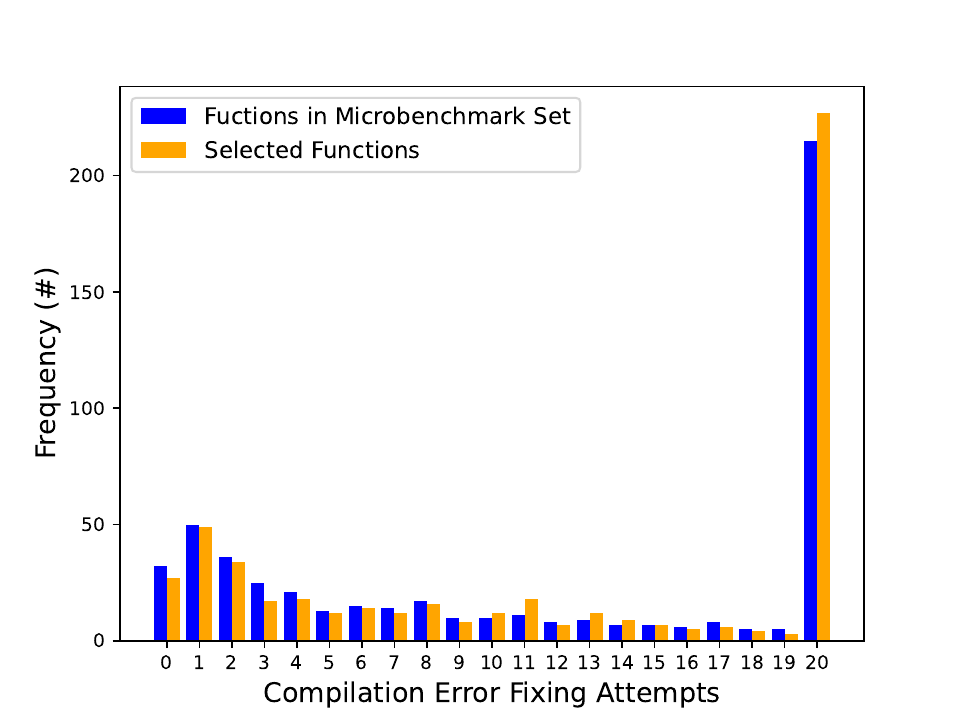}} \\[0.0cm]
    
    \subfloat[qwen2.5-coder:7b]{\includegraphics[width=0.25\textwidth]{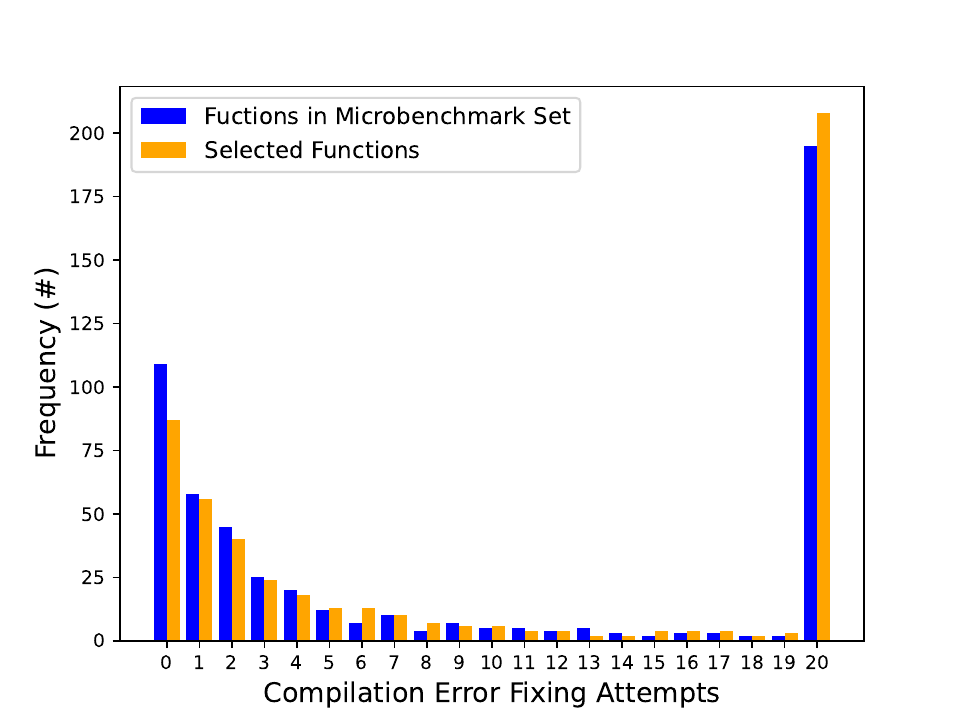}}%
    \subfloat[qwen2.5-coder:14b]{\includegraphics[width=0.25\textwidth]{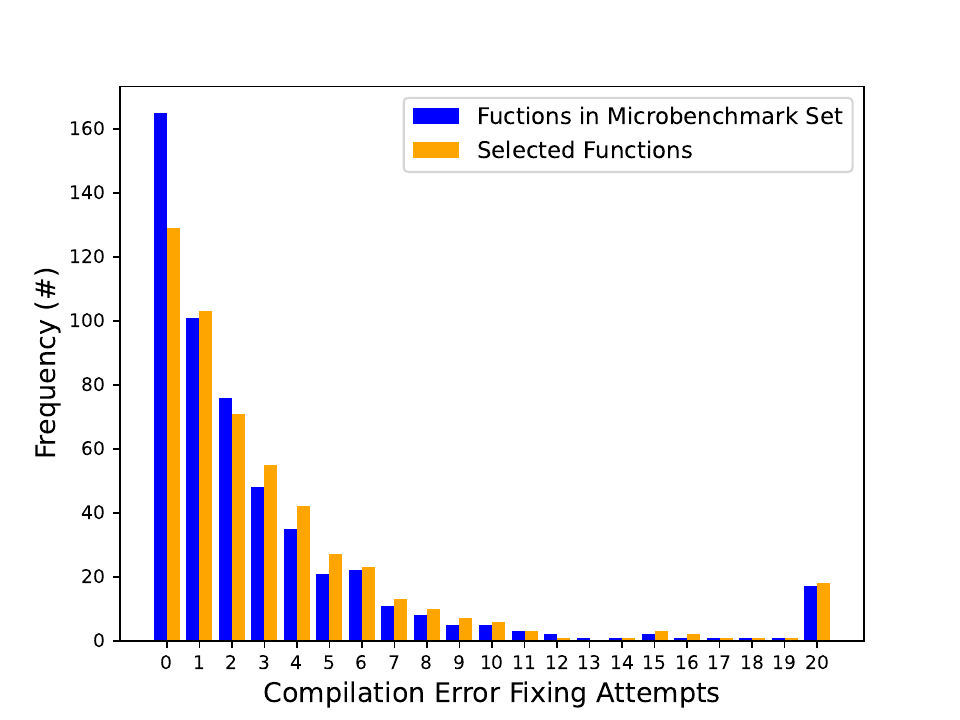}}%
    \subfloat[qwen2.5-coder:32b]{\includegraphics[width=0.25\textwidth]{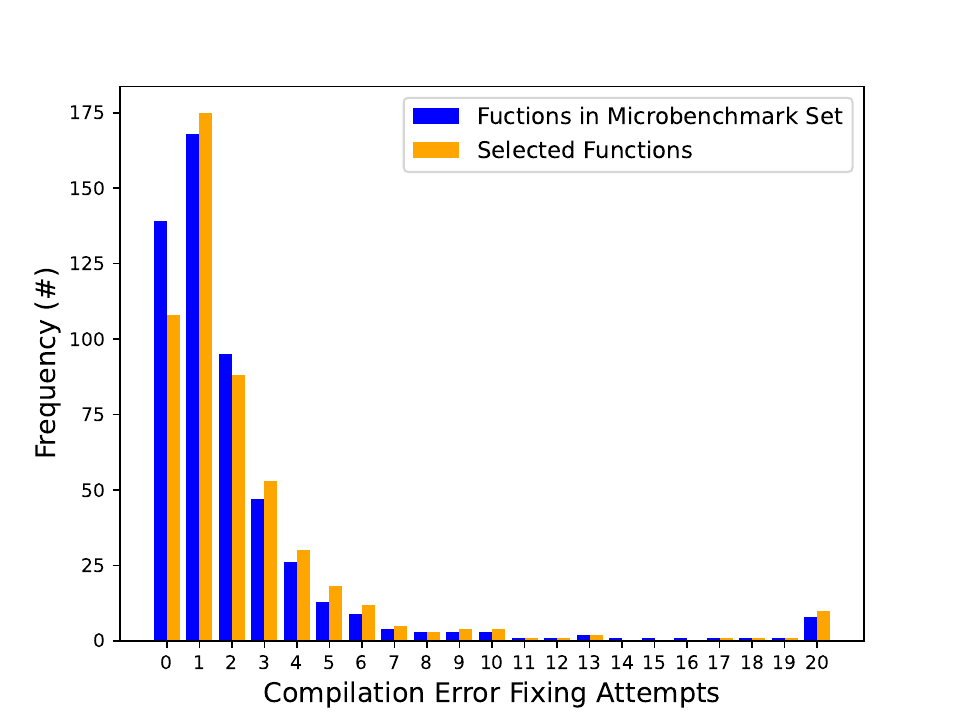}}

    \caption{Compilation error fixing attempt distribution of selected and microbenchmark sets for 9 LLMs.}
    \label{fig:cross_llm_dist}
\end{figure*}

To select best hyperparameters, we look for minimizing the relative difference score. The combination (9, 0.166), placed at the index of 883, which is marked by the red dashed line in \autoref{fig:squared_diff}, has the lowest relative difference score of 11.2\%. Thus, we pick the number of partition and ratio of sampling parameters as 9 and 0.166, respectively.

\subsection{Cross-LLM Evaluation of Selected Functions on Microbenchmark Set}
\label{sec:cross_evaluation}
In this subsection, we perform function selection using the LLM and the hyperparameters previously chosen. Then, we evaluate the selected function set on all 9 LLMs to demonstrate that the selections are generalized to other LLMs.

\paragraph{Methodology.} We perform function selection from microbenchmark set by setting the LLM, the number of partition per dimension, and the ratio of sampling per bin as \texttt{qwen2.5-coder:32b}, 9, and 0.166 respectively. To evaluate the selected functions, we use the compilation error fixing attempt from the transpilation process as in~\S\ref{sec:tuning_hyperparameters}. First, we obtain the distribution of compilation error fixing attempts for both the selected set and the microbencmark set. We normalize the distributions of the microbenchmark set by the ratio of the two sets. Then, we calculate a relative difference score using the formula shown in~\S\ref{sec:tuning_hyperparameters}.

As presented in \S\ref{sec:llm_selection}, the microbencmark set is transpiled using 9 different LLMs. Thus, each of the LLMs has their own compilation error fixing attempt feedback. We calculate the relative difference score for 9 LLMs using their own distributions for selected function set and microbenchmark set. 
Using their individual feedback allows us to justify that the functions selected using \texttt{qwen2.5-coder:32b} are also representative for the other LLMs.

\paragraph{Results.} In \autoref{fig:llm_diff}, we present the relative difference scores for 9 LLMs. For more details, we share the expected and observed distributions for each LLM in \autoref{fig:cross_llm_dist}. Furthermore, the difference score for \texttt{qwen2.5-coder:32b} is different from that presented in~\S\ref{sec:tuning_hyperparameters}, since this experiment is performed on the microbenchmark set. However, the difference scores of the LLMs are comparable to each other, since we use the same set of hyperparameters for all of them.

\begin{table*}[h!]
\centering
\footnotesize
\setlength\tabcolsep{9pt}
\caption{The detailed information of the programs in \nameee{}.}
\label{tab:selectedset_info}
\begin{tabular}{lrr||lrr||lrr}
\toprule
\textbf{Name} & \textbf{Function (\#)} & \textbf{SloC (\#)} & \textbf{Name} & \textbf{Function (\#)} & \textbf{SloC (\#)} & \textbf{Name} & \textbf{Function (\#)} & \textbf{SloC (\#)} \\
\hline
\hline
transcoder-set & 714 & 3,457 &  mtools-4.0.43 & 42 & 463 &      lodepng & 10 & 211 \\ 
libxml2 & 392 & 5,952 &         dap-3.10 & 38 & 1,568 &         diffutils-3.10 & 10 & 186 \\ 
json.h & 135 & 4,641 &  cflow-1.7 & 33 & 383 &  pexec-1.0rc8 & 9 & 90 \\ 
gprolog-1.5.0 & 125 & 1,358 &   bc-1.07.1 & 32 & 981 &  gzip-1.12 & 7 & 221 \\ 
nettle-3.9.1 & 112 & 1,612 &    patch-2.7.6 & 31 & 410 &        lil & 7 & 58 \\ 
libosip2-5.3.1 & 105 & 1,466 &  make-4.4.1 & 29 & 1,537 &       buffer-0.4.0 & 7 & 56 \\ 
tmux & 97 & 1,116 &     rcs-5.10.1 & 29 & 422 &         bzip2 & 5 & 272 \\ 
tulipindicators-0.9.1 & 93 & 1,071 &    findutils-4.9.0 & 27 & 1,141 &  libzahl-1.0 & 5 & 110 \\ 
tar-1.34 & 85 & 1,489 &         binn-3.0 & 26 & 275 &   libcsv & 5 & 29 \\ 
optipng-0.7.8 & 68 & 1,535 &    units-2.22 & 23 & 786 &         snudown & 4 & 45 \\ 
less-633 & 67 & 988 &   pth-2.0.7 & 23 & 374 &  genann & 3 & 84 \\ 
nano-7.2 & 59 & 868 &   cpio-2.14 & 22 & 293 &  quadtree-0.1.0 & 3 & 24 \\ 
gawk-5.2.2 & 56 & 3,182 &       json-c & 22 & 227 &     xzoom & 2 & 380 \\ 
mcsim-6.2.0 & 46 & 759 &        brotli-1.0.9 & 14 & 516 &       which-2.21 & 2 & 222 \\ 
wget-1.21.4 & 46 & 719 &        grep-3.11 & 12 & 871 &  sed-4.9 & 2 & 38 \\ 
uucp-1.07 & 43 & 2,406 &        ed-1.19 & 12 & 126 &    urlparser & 2 & 27 \\ 
heman & 43 & 599 &      indent-2.2.13 & 11 & 382 &      avl & 2 & 16 \\ 
screen-4.9.0 & 42 & 2,066 &     hello-2.12.1 & 11 & 125 &       robotfindskitten & 2 & 7 \\ 
tinycc & 42 & 656 &     enscript-1.6.6 & 10 & 1,155 &   ht & 1 & 14 \\ 
\bottomrule
\end{tabular}
\end{table*}

The five models, including three models of \texttt{qwen2.5-coder}, \texttt{llama3.1:8b}, and \texttt{codestral:22b}, achieve reasonably close difference scores. The difference scores of \texttt{codegeex4:9b} and \texttt{gemma2:9b} are slightly higher than the previously mentioned 5 LLMs while \texttt{llama3.2:3b} and \texttt{mistral:7b} are lower than them. Note that the poor performance of \texttt{llama3.2:3b} and \texttt{mistral:7b} in transpilation leads to uneven distributions with some accumulations as presented in \autoref{fig:cross_llm_dist}, which is the main reason for lower difference scores than \texttt{qwen2.5-coder:32b}. Consequently, even though function selection is performed using the transpilation output of \texttt{qwen2.5-coder:32b}, the relative difference scores of other LLMs show that the selected functions are representative for all LLMs. Thus, employing a specific LLM as part of the selection process for the C-to-Rust transpilation evaluation dataset does not pose a threat to the validity of our study.~\looseness=-1

\subsection{\nameee{}: A Minimized Representative Set}
\label{sec:final_selection}
In this subsection, we share details of \nameee{}. After finalizing the tuning on the LLM and hyperparameters, we perform the selection process on the large dataset that contains \fullset{} number of functions. Our selection process identifies \selectedset{} number of functions as the representative set of the large dataset, which forms \nameee{}. In~\autoref{tab:selectedset_info}, we present the details of \nameee{}. In the function column, we share the total number of individual functions for each program existing in \nameee{}. In the SLoC column, we share the total number of source lines of code for the functions of each program.

By our selection process, the number of functions is reduced from \fullset{} to \selectedset{} by 81.3\% decrease. The transpilation of the programs from the large dataset took 246 hours in total, while the representative set, \nameee{}, took 52 hours in total with a reduction of 78.9\% in time. Moreover, the total source line of code (SLoC) is reduced from 195,926 to 50,150 by 74.4\% decrease.

\section{Related Work}
\paragraph{Transpilation.} Initial attempts have been made to translate a C program into Rust. c2rust is the most popular rule-based transpilation tool, which produces compilable Rust code while preserving semantics ~\cite{immunantc2rust}. However, Rust code produced by c2rust is covered in unsafe blocks and is not idiomatic. Previous works focus on improving the c2rust transpilation output, with the aim of reducing unsafe usages by getting feedback from the Rust compiler~\cite{emre2021translating} and using ownership analysis to convert pointers~\cite{zhang2023ownership}. Another work proposes a technique to improve the c2rust output by replacing the output parameters with Rust's algebraic data types~\cite{hong2024don}.

Although several attempts have been made, rule-based transpilation is fundamentally limited in automated C-to-Rust transpilation. With recent advancements in code tasks using LLMs, the new line of work aims to solve C-to-Rust transpilation using LLMs instead of rule-based approaches. 
VERT combines WebAssembly and LLMs to generate correct and readable Rust code transpiling from various languages, including C~\cite{yang2024vert}. Moreover, another work aims to overcome the limited context windows of LLMs by segmenting input C code and transpiling smaller units into Rust code~\cite{shiraishi2024context}.

\paragraph{Dataset Reduction for Machine Learning.} For machine learning classifiers, the well-established line of research focuses on instance and coreset selection to reduce the training set.
Instance selection studies focus on choosing a subset of representative instances from the original dataset that maintains the overall structure of the data~\cite{olvera2010review}. They aim to preserve important patterns, such as class distributions and feature relationships, while removing redundant or irrelevant data points. Coreset selection studies aim to identify a smaller, weighted subset of data that can approximate the distribution of the entire dataset with minimal loss of accuracy~\cite{har2005smaller, tsang2005core, bachem2017practical, novikov2021dataset}.

Recent work has shifted the focus to coreset selection for neural networks~\cite{sener2017active, lee2024coreset, song2025leave}. A previous study redefines active learning for convolutional neural networks (CNNs) as coreset selection, showing that selecting a subset of data based on geometric properties improves performance over traditional active learning heuristics~\cite{sener2017active}. Another work introduces coreset selection for Object Detection (CSOD) and achieves improved performance over random selection in object detection tasks~\cite{lee2024coreset}. Lastly, a recent work introduces the SubPIE algorithm for coreset selection, which optimizes the coverage radii of coreset elements using entropy-based methods and discrete coordinate descent~\cite{song2025leave}.

\section{Conclusion}
Dealing with memory safety vulnerabilities for more than two decades showed that it is a never ending problem. Thus, migrating from the memory-unsafe C language to a more controlled fundamentally memory-safe Rust language, is seen as the promising solution that can critically reduce memory-safety related exploitations. Since it is nearly impossible to manually migrate all C codebases, it is necessary to have an automated C-to-Rust transpilation framework. However, as in other program analysis tasks, one needs to have a dataset to evaluate proposed solutions. Such dataset should be minimized to reduce resource consumption while covering representative samples.
In this work, applying our selection method, we select representative functions and reduce the large dataset, obtained from previous works, from \fullset{} number of functions to \selectedset{} number of functions that form \nameee{}. We propose \nameee{} to be used in the evaluation of the C-to-Rust transpilation in future work.

\bibliographystyle{ACM-Reference-Format}
\bibliography{paper}


\begin{thebibliography}{27}


\ifx \showCODEN    \undefined \def \showCODEN     #1{\unskip}     \fi
\ifx \showDOI      \undefined \def \showDOI       #1{#1}\fi
\ifx \showISBNx    \undefined \def \showISBNx     #1{\unskip}     \fi
\ifx \showISBNxiii \undefined \def \showISBNxiii  #1{\unskip}     \fi
\ifx \showISSN     \undefined \def \showISSN      #1{\unskip}     \fi
\ifx \showLCCN     \undefined \def \showLCCN      #1{\unskip}     \fi
\ifx \shownote     \undefined \def \shownote      #1{#1}          \fi
\ifx \showarticletitle \undefined \def \showarticletitle #1{#1}   \fi
\ifx \showURL      \undefined \def \showURL       {\relax}        \fi
\providecommand\bibfield[2]{#2}
\providecommand\bibinfo[2]{#2}
\providecommand\natexlab[1]{#1}
\providecommand\showeprint[2][]{arXiv:#2}

\bibitem[cis(2025)]%
        {cisareport2023}
 \bibinfo{year}{2025}\natexlab{}.
\newblock \bibinfo{title}{The Urgent Need for Memory Safety in Software Products}.
\newblock
\newblock
\urldef\tempurl%
\url{https://www.cisa.gov/news-events/news/urgent-need-memory-safety-software-products}
\showURL{%
\tempurl}


\bibitem[whi(2025)]%
        {whitehousereport2024}
 \bibinfo{year}{2025}\natexlab{}.
\newblock \bibinfo{title}{Fact Sheet: {ONCD} Report Calls for Adoption of Memory Safe Programming Languages and Addressing the Hard Research Problem of Software Measurability}.
\newblock
\newblock
\urldef\tempurl%
\url{https://bidenwhitehouse.archives.gov/oncd/briefing-room/2024/02/26/memory-safety-fact-sheet/}
\showURL{%
\tempurl}


\bibitem[dar(2025)]%
        {darpatractor}
 \bibinfo{year}{2025}\natexlab{}.
\newblock \bibinfo{title}{TRACTOR: Translating All C to Rust}.
\newblock
\newblock
\urldef\tempurl%
\url{https://www.darpa.mil/research/programs/translating-all-c-to-rust}
\showURL{%
\tempurl}


\bibitem[Emre et~al\mbox{.}(2021)]%
        {emre2021translating}
\bibfield{author}{\bibinfo{person}{Mehmet Emre}, \bibinfo{person}{Ryan Schroeder}, \bibinfo{person}{Kyle Dewey}, {and} \bibinfo{person}{Ben Hardekopf}.} \bibinfo{year}{2021}\natexlab{}.
\newblock \showarticletitle{Translating C to safer Rust}.
\newblock \bibinfo{journal}{\emph{Proceedings of the ACM on Programming Languages}} \bibinfo{volume}{5}, \bibinfo{number}{OOPSLA} (\bibinfo{year}{2021}), \bibinfo{pages}{1--29}.
\newblock


\bibitem[Zhang et~al\mbox{.}(2023)]%
        {zhang2023ownership}
\bibfield{author}{\bibinfo{person}{Hanliang Zhang}, \bibinfo{person}{Cristina David}, \bibinfo{person}{Yijun Yu}, {and} \bibinfo{person}{Meng Wang}.} \bibinfo{year}{2023}\natexlab{}.
\newblock \showarticletitle{Ownership guided C to Rust translation}. In \bibinfo{booktitle}{\emph{International Conference on Computer Aided Verification}}. Springer, \bibinfo{pages}{459--482}.
\newblock


\bibitem[Yang et~al\mbox{.}(2024)]%
        {yang2024vert}
\bibfield{author}{\bibinfo{person}{Aidan~ZH Yang}, \bibinfo{person}{Yoshiki Takashima}, \bibinfo{person}{Brandon Paulsen}, \bibinfo{person}{Josiah Dodds}, {and} \bibinfo{person}{Daniel Kroening}.} \bibinfo{year}{2024}\natexlab{}.
\newblock \showarticletitle{VERT: Verified equivalent rust transpilation with large language models as few-shot learners}.
\newblock \bibinfo{journal}{\emph{arXiv preprint arXiv:2404.18852}} (\bibinfo{year}{2024}).
\newblock


\bibitem[Hong and Ryu(2024)]%
        {hong2024don}
\bibfield{author}{\bibinfo{person}{Jaemin Hong} {and} \bibinfo{person}{Sukyoung Ryu}.} \bibinfo{year}{2024}\natexlab{}.
\newblock \showarticletitle{Don’t Write, but Return: Replacing Output Parameters with Algebraic Data Types in C-to-Rust Translation}.
\newblock \bibinfo{journal}{\emph{Proceedings of the ACM on Programming Languages}} \bibinfo{volume}{8}, \bibinfo{number}{PLDI} (\bibinfo{year}{2024}), \bibinfo{pages}{716--740}.
\newblock


\bibitem[Shiraishi and Shinagawa(2024)]%
        {shiraishi2024context}
\bibfield{author}{\bibinfo{person}{Momoko Shiraishi} {and} \bibinfo{person}{Takahiro Shinagawa}.} \bibinfo{year}{2024}\natexlab{}.
\newblock \showarticletitle{Context-aware Code Segmentation for C-to-Rust Translation using Large Language Models}.
\newblock \bibinfo{journal}{\emph{arXiv preprint arXiv:2409.10506}} (\bibinfo{year}{2024}).
\newblock


\bibitem[Dolan-Gavitt et~al\mbox{.}(2016)]%
        {dolan2016lava}
\bibfield{author}{\bibinfo{person}{Brendan Dolan-Gavitt}, \bibinfo{person}{Patrick Hulin}, \bibinfo{person}{Engin Kirda}, \bibinfo{person}{Tim Leek}, \bibinfo{person}{Andrea Mambretti}, \bibinfo{person}{Wil Robertson}, \bibinfo{person}{Frederick Ulrich}, {and} \bibinfo{person}{Ryan Whelan}.} \bibinfo{year}{2016}\natexlab{}.
\newblock \showarticletitle{Lava: Large-scale automated vulnerability addition}. In \bibinfo{booktitle}{\emph{2016 IEEE symposium on security and privacy (SP)}}. IEEE, \bibinfo{pages}{110--121}.
\newblock


\bibitem[Hazimeh et~al\mbox{.}(2020)]%
        {hazimeh2020magma}
\bibfield{author}{\bibinfo{person}{Ahmad Hazimeh}, \bibinfo{person}{Adrian Herrera}, {and} \bibinfo{person}{Mathias Payer}.} \bibinfo{year}{2020}\natexlab{}.
\newblock \showarticletitle{Magma: A ground-truth fuzzing benchmark}.
\newblock \bibinfo{journal}{\emph{Proceedings of the ACM on Measurement and Analysis of Computing Systems}} \bibinfo{volume}{4}, \bibinfo{number}{3} (\bibinfo{year}{2020}), \bibinfo{pages}{1--29}.
\newblock


\bibitem[Bachem et~al\mbox{.}(2017)]%
        {bachem2017practical}
\bibfield{author}{\bibinfo{person}{Olivier Bachem}, \bibinfo{person}{Mario Lucic}, {and} \bibinfo{person}{Andreas Krause}.} \bibinfo{year}{2017}\natexlab{}.
\newblock \showarticletitle{Practical coreset constructions for machine learning}.
\newblock \bibinfo{journal}{\emph{arXiv preprint arXiv:1703.06476}} (\bibinfo{year}{2017}).
\newblock


\bibitem[Sener and Savarese(2017)]%
        {sener2017active}
\bibfield{author}{\bibinfo{person}{Ozan Sener} {and} \bibinfo{person}{Silvio Savarese}.} \bibinfo{year}{2017}\natexlab{}.
\newblock \showarticletitle{Active learning for convolutional neural networks: A core-set approach}.
\newblock \bibinfo{journal}{\emph{arXiv preprint arXiv:1708.00489}} (\bibinfo{year}{2017}).
\newblock


\bibitem[Novikov et~al\mbox{.}(2021)]%
        {novikov2021dataset}
\bibfield{author}{\bibinfo{person}{Georgii Novikov}, \bibinfo{person}{Maxim Panov}, {and} \bibinfo{person}{Ivan Oseledets}.} \bibinfo{year}{2021}\natexlab{}.
\newblock \showarticletitle{Dataset Reduction via Bias-Variance Minimization}. In \bibinfo{booktitle}{\emph{2021 5th Scientific School Dynamics of Complex Networks and their Applications (DCNA)}}. IEEE, \bibinfo{pages}{143--146}.
\newblock


\bibitem[Lee et~al\mbox{.}(2024)]%
        {lee2024coreset}
\bibfield{author}{\bibinfo{person}{Hojun Lee}, \bibinfo{person}{Suyoung Kim}, \bibinfo{person}{Junhoo Lee}, \bibinfo{person}{Jaeyoung Yoo}, {and} \bibinfo{person}{Nojun Kwak}.} \bibinfo{year}{2024}\natexlab{}.
\newblock \showarticletitle{Coreset selection for object detection}. In \bibinfo{booktitle}{\emph{Proceedings of the IEEE/CVF Conference on Computer Vision and Pattern Recognition}}. \bibinfo{pages}{7682--7691}.
\newblock


\bibitem[Song et~al\mbox{.}(2025)]%
        {song2025leave}
\bibfield{author}{\bibinfo{person}{Haohao Song}, \bibinfo{person}{Qiao Xiang}, {and} \bibinfo{person}{Jiwu Shu}.} \bibinfo{year}{2025}\natexlab{}.
\newblock \showarticletitle{Leave No Stone Unturned: Optimizing Subpattern Information Entropy for Coreset Selection}. In \bibinfo{booktitle}{\emph{ICASSP 2025-2025 IEEE International Conference on Acoustics, Speech and Signal Processing (ICASSP)}}. IEEE, \bibinfo{pages}{1--5}.
\newblock


\bibitem[Coleman et~al\mbox{.}(1994)]%
        {coleman1994using}
\bibfield{author}{\bibinfo{person}{Don Coleman}, \bibinfo{person}{Dan Ash}, \bibinfo{person}{Bruce Lowther}, {and} \bibinfo{person}{Paul Oman}.} \bibinfo{year}{1994}\natexlab{}.
\newblock \showarticletitle{Using metrics to evaluate software system maintainability}.
\newblock \bibinfo{journal}{\emph{Computer}} \bibinfo{volume}{27}, \bibinfo{number}{8} (\bibinfo{year}{1994}), \bibinfo{pages}{44--49}.
\newblock


\bibitem[McCabe(1976)]%
        {mccabe1976complexity}
\bibfield{author}{\bibinfo{person}{Thomas~J McCabe}.} \bibinfo{year}{1976}\natexlab{}.
\newblock \showarticletitle{A complexity measure}.
\newblock \bibinfo{journal}{\emph{IEEE Transactions on software Engineering}} \bibinfo{number}{4} (\bibinfo{year}{1976}), \bibinfo{pages}{308--320}.
\newblock


\bibitem[Halstead(1977)]%
        {halstead1977elements}
\bibfield{author}{\bibinfo{person}{Maurice~H Halstead}.} \bibinfo{year}{1977}\natexlab{}.
\newblock \bibinfo{booktitle}{\emph{Elements of Software Science (Operating and programming systems series)}}.
\newblock \bibinfo{publisher}{Elsevier Science Inc.}
\newblock


\bibitem[immunant(2025)]%
        {immunantc2rust}
\bibfield{author}{\bibinfo{person}{immunant}.} \bibinfo{year}{2025}\natexlab{}.
\newblock \bibinfo{title}{c2rust}.
\newblock
\newblock
\urldef\tempurl%
\url{https://github.com/immunant/c2rust}
\showURL{%
\tempurl}


\bibitem[Song et~al\mbox{.}(2019)]%
        {song2019sok}
\bibfield{author}{\bibinfo{person}{Dokyung Song}, \bibinfo{person}{Julian Lettner}, \bibinfo{person}{Prabhu Rajasekaran}, \bibinfo{person}{Yeoul Na}, \bibinfo{person}{Stijn Volckaert}, \bibinfo{person}{Per Larsen}, {and} \bibinfo{person}{Michael Franz}.} \bibinfo{year}{2019}\natexlab{}.
\newblock \showarticletitle{SoK: Sanitizing for security}. In \bibinfo{booktitle}{\emph{2019 IEEE Symposium on Security and Privacy (SP)}}. IEEE, \bibinfo{pages}{1275--1295}.
\newblock


\bibitem[Oman and Hagemeister(1992)]%
        {oman1992metrics}
\bibfield{author}{\bibinfo{person}{Paul Oman} {and} \bibinfo{person}{Jack Hagemeister}.} \bibinfo{year}{1992}\natexlab{}.
\newblock \showarticletitle{Metrics for assessing a software system's maintainability}. In \bibinfo{booktitle}{\emph{Proceedings Conference on Software Maintenance 1992}}. IEEE Computer Society, \bibinfo{pages}{337--338}.
\newblock


\bibitem[Jarod42(2025)]%
        {Jarod42}
\bibfield{author}{\bibinfo{person}{Jarod42}.} \bibinfo{year}{2025}\natexlab{}.
\newblock \bibinfo{title}{ccccc}.
\newblock
\newblock
\urldef\tempurl%
\url{https://github.com/Jarod42/ccccc}
\showURL{%
\tempurl}


\bibitem[Ardito et~al\mbox{.}(2020)]%
        {ardito2020rust}
\bibfield{author}{\bibinfo{person}{Luca Ardito}, \bibinfo{person}{Luca Barbato}, \bibinfo{person}{Marco Castelluccio}, \bibinfo{person}{Riccardo Coppola}, \bibinfo{person}{Calixte Denizet}, \bibinfo{person}{Sylvestre Ledru}, {and} \bibinfo{person}{Michele Valsesia}.} \bibinfo{year}{2020}\natexlab{}.
\newblock \showarticletitle{rust-code-analysis: A Rust library to analyze and extract maintainability information from source codes}.
\newblock \bibinfo{journal}{\emph{SoftwareX}}  \bibinfo{volume}{12} (\bibinfo{year}{2020}), \bibinfo{pages}{100635}.
\newblock


\bibitem[immunant(2025)]%
        {immunantc2rustexamples}
\bibfield{author}{\bibinfo{person}{immunant}.} \bibinfo{year}{2025}\natexlab{}.
\newblock \bibinfo{title}{c2rust}.
\newblock
\newblock
\urldef\tempurl%
\url{https://github.com/immunant/c2rust/tree/master/examples}
\showURL{%
\tempurl}


\bibitem[Olvera-L{\'o}pez et~al\mbox{.}(2010)]%
        {olvera2010review}
\bibfield{author}{\bibinfo{person}{J~Arturo Olvera-L{\'o}pez}, \bibinfo{person}{J~Ariel Carrasco-Ochoa}, \bibinfo{person}{J~Francisco Mart{\'\i}nez-Trinidad}, {and} \bibinfo{person}{Josef Kittler}.} \bibinfo{year}{2010}\natexlab{}.
\newblock \showarticletitle{A review of instance selection methods}.
\newblock \bibinfo{journal}{\emph{Artificial Intelligence Review}}  \bibinfo{volume}{34} (\bibinfo{year}{2010}), \bibinfo{pages}{133--143}.
\newblock


\bibitem[Har-Peled and Kushal(2005)]%
        {har2005smaller}
\bibfield{author}{\bibinfo{person}{Sariel Har-Peled} {and} \bibinfo{person}{Akash Kushal}.} \bibinfo{year}{2005}\natexlab{}.
\newblock \showarticletitle{Smaller coresets for k-median and k-means clustering}. In \bibinfo{booktitle}{\emph{Proceedings of the twenty-first annual symposium on Computational geometry}}. \bibinfo{pages}{126--134}.
\newblock


\bibitem[Tsang et~al\mbox{.}(2005)]%
        {tsang2005core}
\bibfield{author}{\bibinfo{person}{Ivor~W Tsang}, \bibinfo{person}{James~T Kwok}, \bibinfo{person}{Nello Cristianini}, {et~al\mbox{.}}} \bibinfo{year}{2005}\natexlab{}.
\newblock \showarticletitle{Core vector machines: Fast SVM training on very large data sets.}
\newblock \bibinfo{journal}{\emph{Journal of machine Learning research}} \bibinfo{volume}{6}, \bibinfo{number}{4} (\bibinfo{year}{2005}).
\newblock


\end{thebibliography}

\appendix
\section{Implementation Details}
\subsection{Transpilation of C Functions into Rust}
\label{appendix:transpilation_implementation}

\begin{figure}[h]
\begin{tcolorbox}[colframe=black!25, colback=yellow!10, coltitle=blue!20!black]
Behave like you are an expert of C and Rust. Behave like you are a translator from C language to Rust language. Can you translate C code given above into Rust code?

Do not explain the code to me! Only return Rust code correspoding to the given C code.
\\

Follow these intructions strictly in translation:

(1) Do not add any extra error handling,

(2) Do not merge functions,

(3) Do not change variable names,

(4) Use no\_mangle for each function,

(5) Make each function public,

(6) Translate the standard C library function calls by placing a decoy function call (leave the decoy function body empty if possible) with the same name, and

(7) Only return a Rust code and nothing else!
\end{tcolorbox}
\caption{The instructions given to LLM for initial transpilation.\label{fig:transpile_instruction}}
\end{figure}

In \autoref{fig:transpile_instruction}, we present the instructions sent to LLM along with the input C function for the initial transpilation from the C to Rust. 

\begin{figure}[h]
\begin{tcolorbox}[colframe=black!25, colback=yellow!10, coltitle=blue!20!black]
When attempted to compile the recently generated rust code, I obtained the compilation errors given above. Fix those errors and only return the modified Rust code. Do not explain the code or changes to me!
\end{tcolorbox}
\caption{The instructions given to LLM for fixing compilation errors.\label{fig:fixing_instruction}}
\end{figure}

In \autoref{fig:fixing_instruction}, we present the instructions sent to LLM along with the compilation errors to fix the compilation errors previously obtained from the compilation attempt of transpiled Rust code.

\end{document}